\documentclass[lettersize,journal]{IEEEtran}
\usepackage{graphicx} 
\usepackage{amsmath,amsfonts,amssymb}
\usepackage{algorithmic}
\usepackage{indentfirst} 
\usepackage{algorithm}
\usepackage{array}
\usepackage{eurosym}
\usepackage{bm}
\usepackage{gensymb}
\usepackage{caption}
\usepackage{lettrine}
\usepackage{float}
\usepackage{textcomp}
\usepackage{stfloats}
\usepackage{ulem}
\usepackage{url}
\usepackage{enumerate}
\usepackage{nccmath} 
\usepackage{verbatim}
\usepackage{cite}
\usepackage[caption=false,font=footnotesize]{subfig}
\usepackage{textcomp}
\usepackage{xcolor}
\usepackage{bm}
\usepackage{subcaption}
\makeatletter \renewcommand{\maketag@@@}[1]{\hbox{\m@th\normalsize\normalfont#1}}
\usepackage{subfloat}
\usepackage{balance}
\usepackage{autobreak}
\usepackage{url}
\usepackage{cite}
\usepackage {epstopdf}
\allowdisplaybreaks[4]
\usepackage{orcidlink} %
\hypersetup{hidelinks} 

\title{\huge Robust Secure Beamforming for Movable Antenna Enhanced Integrated Sensing and Communications}
\author{Yuan Chen, Ning Wei, \IEEEmembership{Member,~IEEE}, Ahmad Bazzi, \IEEEmembership{Senior Member,~IEEE},\\
Xiangyu Dong, Ran Yang, You Li, and Yue Xiu, \IEEEmembership{Member,~IEEE}
\thanks{%
    This work was supported in part by Mobile Information Networks-National Science and Technology Major Project under Grant 2025ZD1302000, in part by Beijing Municipal Science and Technology Program(No. Z24110100120000), in part by China Mobile Communications Group Co., Ltd. \textit{(Corresponding author: Ning Wei.)}
    
    Yuan Chen, Ning Wei, Xiangyu Dong, Ran Yang, and Yue Xiu are with the National Key Laboratory of Wireless Communications, University of Electronic Science and Technology of China, Chengdu 611731, China (e-mail: wn@uestc.edu.cn).%

    Ahmad Bazzi is with New York University (NYU) Abu Dhabi, and NYU WIRELESS, NYU Tandon School of Engineering, Brooklyn, 11201, NY, USA (e-mail: ahmad.bazzi@nyu.edu).

    You Li is with the Southwest China Research Institute of Electronic
    Equipment (SWIEE), China (e-mail:liyou1992@163.com).

}}
\IEEEaftertitletext{\vspace{-12mm}}
\date{November 2025}
\usepackage{titlesec}

\titlespacing*{\section}
{0pt}      
{0.3ex }   
{0.1ex }   

\titlespacing*{\subsection}
{0pt}
{0.3ex }
{0.1ex }

\titlespacing*{\subsubsection}
{0pt}
{0.3ex }
{0.1ex }
\begin{document}	
\maketitle

\begin{abstract}
In this letter, we investigate robust beamforming design for a movable antenna (MA)-enhanced secure integrated sensing and communications (ISAC) system with imperfect eavesdropping channel state information (CSI). To improve radar sensing performance, we formulate a radar signal-to-interference-plus-noise ratio (SINR) maximization problem by jointly optimizing the transmit beamforming and antenna position while ensuring communication data security. However, the resulting optimization problem is inherently intractable due to the nonlinear mapping from antenna positions to channel coefficients, as well as the eavesdropper (Eve) channel uncertainty. To handle these challenges, we propose a block coordinate descent (BCD)-based algorithm incorporating successive convex approximation (SCA) and fractional programming (FP) techniques. Simulation results show that our proposed algorithm exhibits fast convergence and achieves a significant improvement in the radar SINR while guaranteeing communication security.
\end{abstract}
\section{Introduction}
\IEEEPARstart{I}{ntegrated} sensing and communications (ISAC) has been identified as one of the key usage scenarios for International Mobile Telecommunications toward 2030 and beyond (IMT-2030) \cite{baduge2026}. By enabling communication and sensing functions to share the same spectrum, hardware, and signaling resources, ISAC is expected to support environment-aware wireless services such as intelligent transportation, autonomous systems, and industrial automation \cite{bazzi2026}. However, as the unified waveform is used for target probing, confidential information becomes vulnerable  to malicious eavesdroppers. The open broadcast nature of the wireless environment further compounds this vulnerability. Therefore, developing a  security signaling scheme becomes an important and practically relevant issue in ISAC systems \cite{wei2022toward}.

Recently, many security measures have been developed, such as artificial noise (AN) \cite{zhang2025AN}, dynamic resource allocation \cite{xu2022resource-allocation},  and covert transmission \cite{hu2024covert}. Although these approaches effectively protect confidential communication data from interception, existing approaches are still based on conventional fixed-position antennas (FPAs), with limited spatial degrees of freedom (DoFs). Specifically, this restriction results in  a reduced ability to shape the MIMO channel, control interference, and enhance spatial resolution, leading to degradation of  both radar and communication performance \cite{ma2025secure-ISAC-ma}.

Recently, movable antenna (MA) has emerged as a promising solution to overcome the limitations of FPAs. Different from conventional arrays, MAs can dynamically adjust antenna locations within a prescribed region, thereby introducing additional spatial DoFs for channel reshaping and interference control. Recent studies have shown the potential of MA technology in boosting dual-task performance in ISAC systems \cite{Xiu1,Xiu2,Xiu3,Xiu4}. For example, antenna-position optimization was performed in \cite{ma2026movable} to enhance the communication performance. In \cite{lyu2024flexible}, the sensing and communication mutual information (MI) was maximized by jointly optimizing antenna position and beamforming vectors. The authors in \cite{hu2024secure} investigated the joint antenna position and beamforming to secure the ISAC system. In \cite{yang2025covertMA}, the authors proposed a covert transmission framework for MA-assisted ISAC systems. Despite the demonstrated efficiency of MAs, these works are typically based on perfect wiretap channel state information  (CSI). In practical ISAC systems, acquiring the perfect eavesdropper (Eve) CSI is highly challenging due to the non-cooperative and passive nature of Eve. Therefore, developing robust beamforming against  Eve channel uncertainty is of great significance for MA-enhanced ISAC systems.

In this letter, we investigate a robust secure beamforming design for an MA-enhanced ISAC system under imperfect Eve CSI. The main contributions are summarized as: 

1) To enhance the sensing performance, we formulate a radar signal-to-interference-plus-noise ratio (SINR) maximization problem for an MA-enhanced ISAC system under imperfect Eve CSI, subject to the communication and security requirements. The optimization is achieved by jointly optimizing the transmit beamforming vectors and transceiver antenna positions. 

2) To solve the highly intractable problem, we develop an iterative algorithm based on fractional programming (FP), block coordinate descent (BCD), and successive convex approximation (SCA) techniques. 

3) Simulation results demonstrate that the proposed design improves the radar SINR, achieves a favorable sensing-security tradeoff, and maintains remarkable robustness against Eve channel uncertainty compared to benchmark schemes.

\section{System Model}
\begin{figure}[t]
    \centering
    \includegraphics[width=0.80\linewidth]{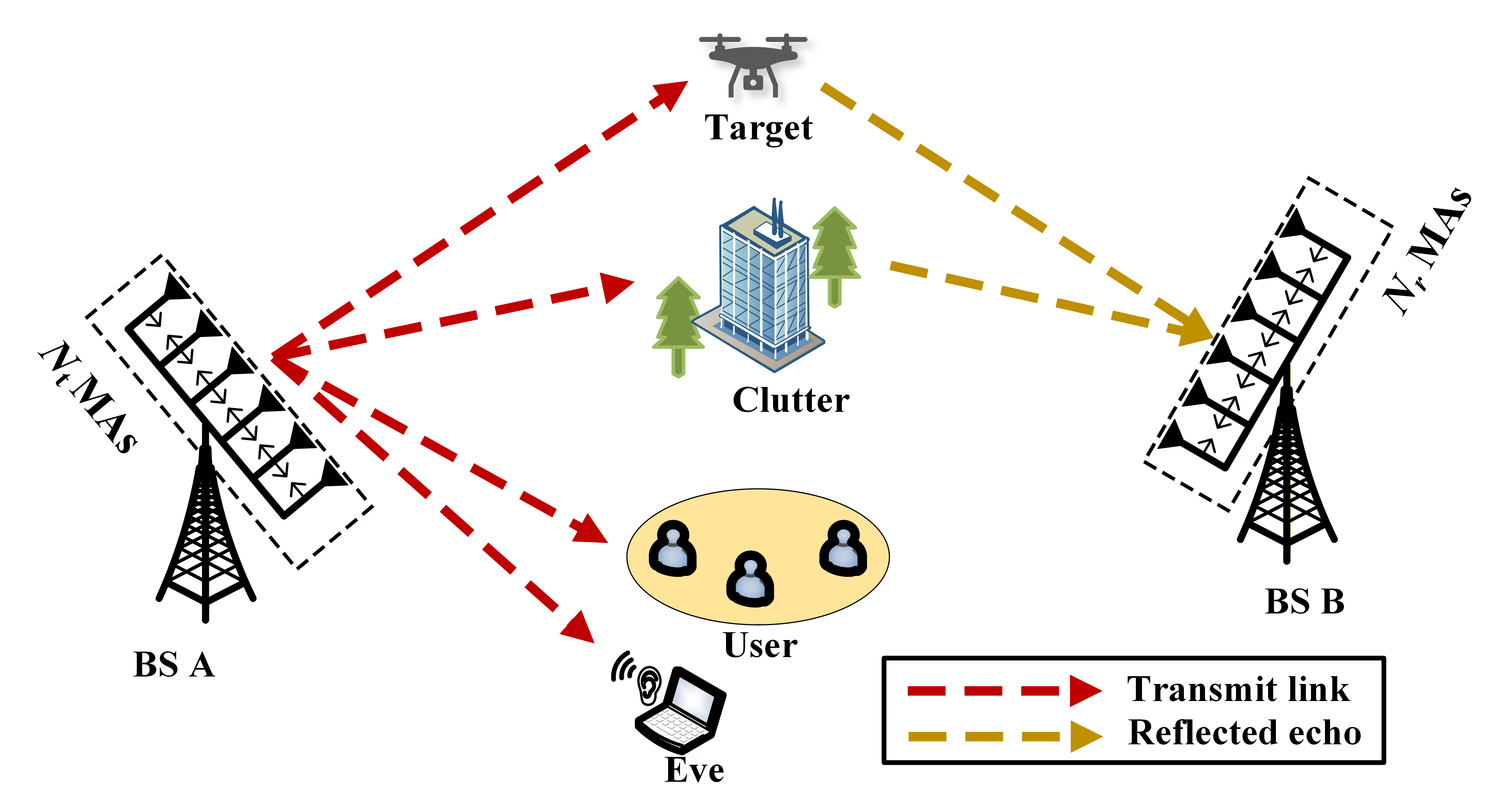}
    \caption{\makebox[\linewidth][l]{MA-enhanced ISAC system.}}
    \label{fig1}
\end{figure}

We consider a bi-static MA-enhanced ISAC system, as shown in Fig.~\ref{fig1}, where both the transmit and receive base stations (BSs) are equipped with $N_t$ and $N_r$ movable antennas arranged as a linear array, respectively. Specifically, the transmit BS, referred to as BS~A, serves $K$ legitimate users in the presence of an Eve. All users and Eve are equipped with a single fixed antenna. BS~A transmits dual-functional signals for communication and sensing, while the receive BS, referred to as BS~B, cooperatively receives the target echo.
	\subsection{Signal Model}
    Let $\bm{x}\in \mathbb{C}^{N_t\times1}$ denote the transmission signal, given by 
	\begin{align}
		\bm{x} =\bm{F}_c \bm{s}_c+\bm{F}_r \bm{s}_r = \bm{F}\bm{s},
	\end{align}
    where $\bm{F}_c = \begin{bmatrix} \bm{f}_{c,1}, \cdots, \bm{f}_{c,K} \end{bmatrix}  \in \mathbb{C}^{N_t \times K}$ and $\bm{F}_r = \begin{bmatrix} \bm{f}_{r,1}, \cdots, \bm{f}_{r,N_t} \end{bmatrix}  \in \mathbb{C}^{N_t \times N_t}$ denote the transmit beamforming matrices for communication and sensing, respectively. The communication and sensing symbols are denoted by $\bm{s}_c\in\mathbb{C}^{K\times1}$ and  $\bm{s}_r\in\mathbb{C}^{N_t\times1}$, respectively. It is assumed that $\bm{s}_c$ and $\bm{s}_r$ are independent complex Gaussian distributed with $\bm{s}_c \sim \mathcal{CN}(0,\bm{I}_K)$ and $\bm{s}_r \sim \mathcal{CN}(0,\bm{I}_{N_t})$.  We further augment the beamforming matrix  as $\bm{F} = [\bm{F}_c,\bm{F}_r]=[\bm{f}_1,\cdots,\bm{f}_{M}] \in \mathbb{C}^{N_t\times M}$ and the transmit symbol vector as $\bm{s}=[\bm{s}_c^T,\bm{s}_r^T]^T=[s_1,\cdots,s_M]^T \in \mathbb{C}^{M\times1}$, where $M=K+N_t$.\par 
\subsection{Communication Model}
    Given that the signal propagation distance is significantly larger than the size of the antenna moving region, the far-field channel model is adopted \cite{zhu2025tutorial}. Consequently, for each channel path component, all antenna elements within the same Tx array experience the same angle of departure (AoD) and the same amplitude of the path gain. Hence, the array response is based on the phase differences of the complex path coefficients among different antenna elements \cite{chowdary2023}. 
   The propagation environment between BS~A and node $\mu\in\{1,\ldots,K,e\}$ indicating the $k$-th user or Eve, is modeled with $L_{\mu}$ paths, where $\theta_{\mu,l}\in[0,\pi]$ denotes the azimuth angle of the $l$-th path observed at BS~A. The transceiver MA position vectors are denoted by $\tilde{\bm{t}}=[t_1,\cdots,t_{N_t}]^T$ and $\tilde{\bm{r}}=[r_1,\cdots,r_{N_r}]^T$, respectively. Consequently, the field response vector (FRV) of the $l$-th propagation path can be expressed as	$
		\bm{a}_{\mu,l}(\tilde{\bm{t}})= [ e^{j \frac{2\pi}{\lambda} t_1 \cos \theta_{\mu,l}}, \cdots, e^{j \frac{2\pi}{\lambda} t_{N_t} \cos \theta_{\mu,l}} ]^T\in\mathbb{C}^{N_t \times1}$, where $\lambda$ denotes the carrier wavelength. The channel $\bm{h}_\mu(\tilde{\bm t})\in \mathbb{C}^{N_t \times 1}$ from BS~A to the $k$-th user or Eve can be expressed as 
\begin{equation}
	\bm{h}_\mu(\tilde{\bm t})=
	\sqrt{\frac{N_t}{L_{\mu}}}\sum_{l=1}^{L_{\mu}}\rho_{\mu,l}\bm{a}_{\mu,l}(\tilde{\bm t}),
\end{equation}
where $\rho_{\mu,l}$ is the complex path gain for the $l$-th path of the user or Eve. Due to the non-cooperative nature of Eve, its channel is assumed to be imperfectly known, and modeled as
$\bm{h}_e(\tilde{\bm{t}})=\hat{\bm h}_e(\tilde{\bm{t}})+\Delta\bm{h}_e, \
	\|\Delta\bm{h}_e\|_2\leq\varepsilon_e$
where $\hat{\bm h}_e(\tilde{\bm{t}})\in \mathbb{C}^{N_t \times 1}$ is the estimated Eve channel and $\varepsilon_e$ denotes the uncertainty radius. In the proposed system, the quasi-static block fading channels are considered, and the received signals at the $k$-th user and Eve are respectively given by
	\begin{equation}
		y_{k} = \bm{h}_{k}(\tilde{\bm{t}})^H \bm{f}_k s_k + \bm{h}_{k}(\tilde{\bm{t}})^H \sum_{u \neq k}^{M} \bm{f}_u s_u  + n_{k},
	\end{equation}
    
    	\begin{equation}
		y_{e,k} = \bm{h}_{e}(\tilde{\bm{t}})^H \bm{f}_k s_k + \bm{h}_{e}(\tilde{\bm{t}})^H \sum_{u \neq k}^{M} \bm{f}_u s_u  + n_{e},
	\end{equation}
where $n_{k}\sim \mathcal{CN}(0, \sigma_{k}^2)$ and $n_{e}\sim \mathcal{CN}(0, \sigma_{e}^2)$ are the additive white Gaussian
noise (AWGN) at the $k$-th user and Eve, respectively. Then, the SINR at the $k$-th user and the eavesdropping SINR at Eve for the $k$-th user are respectively given by
	\begin{equation}
		\Gamma_{k} =    \frac{\bm{h}_k(\tilde{\bm{t}})^H \bm{f}_k {\bm{f}_k}^H \bm{h}_k(\tilde{\bm{t}})}{\sum_{ u \neq k}^{M} \bm{h}_k(\tilde{\bm{t}})^H \bm{f}_u {\bm{f}_u}^H \bm{h}_k(\tilde{\bm{t}}) + \sigma_k^2},
	\end{equation}
    
	\begin{equation}
		\Gamma_{e,k} = 	\frac{\bm{h}_e(\tilde{\bm{t}})^H \bm{f}_k {\bm{f}_k}^H \bm{h}_e(\tilde{\bm{t}})}{\sum_{u\neq k}^{M} \bm{h}_e(\tilde{\bm{t}})^H \bm{f}_u {\bm{f}_u}^H \bm{h}_e(\tilde{\bm{t}}) + \sigma_e^2}.
	\end{equation}		
	\subsection{Radar Model}
	\par 
    We adopt the line-of-sight (LoS) channel model for the sensing channel between the BSs and the target. Let $\theta_{\chi,t}$ and $\theta_{\chi,r}$ denote the azimuth angles from BS~A to target/clutters and from target/clutters to BS~B, respectively. Accordingly, for $\chi \in \{s,1,\cdots,Q\}$ denoting the target or the $q$-th clutter, the transmit and receive steering vectors can be given by $\bm{a}_{\chi,t}(\tilde{\bm{t}}) = [ e^{j\frac{2\pi}{\lambda}t_1 \cos\theta_{\chi,t}}, \cdots, e^{j\frac{2\pi}{\lambda}t_{N_t} \cos\theta_{\chi,t}}]^T\in \mathbb{C}^{N_t \times 1}$ and $\bm{a}_{\chi,r}(\tilde{\bm{r}}) = [ e^{j\frac{2\pi}{\lambda}r_1 \cos\theta_{\chi,r}}, \cdots, e^{j\frac{2\pi}{\lambda}r_{N_r} \cos\theta_{\chi,r}}]^T\in \mathbb{C}^{N_r \times 1} $, respectively. The received echo signal $y_r$ can be expressed as 
	\begin{equation}
		y_r=\alpha_{0}\bm{v}^H\bm{A}_{0}(\tilde{\bm{r}},\tilde{\bm{t}})\bm{x}+\bm{v}^H\sum_{q=1}^Q \alpha_q  \bm{A}_q(\tilde{\bm{r}},\tilde{\bm{t}}) \bm{x}+\bm{v}^H\bm{n}_{r},
	\end{equation}
	where $\bm{v}$ denotes the receiving filter, $\bm{A}_0(\tilde{\bm{r}},\tilde{\bm{t}})\triangleq \bm{a}_{s,r}(\tilde{\bm{r}}) \bm{a}_{s,t}^H (\tilde{\bm{t}})\in \mathbb{C}^{N_r \times N_t}\ \text{and}\ \bm{A}_q(\tilde{\bm{r}},\tilde{\bm{t}})=\bm{a}_{q,r} (\tilde{\bm{r}})\bm{a}_{q,t}^H(\tilde{\bm{t}})\in \mathbb{C}^{N_r \times N_t}$ denote the response matrix for the target and the $q$-th clutter, respectively. Here, $\alpha_0$ and $\alpha_q$ are the complex coefficients that include the radar cross section (RCS) and the cascaded complex gain of the target and the $q$-th clutter,  with $\mathbb{E}[|\alpha_0|^2]=\zeta_0^2$ and $\mathbb{E}[|\alpha_q|^2]=\zeta_q^2$, respectively. Besides, $\bm{n}_{r}\sim \mathcal{CN}(0, \sigma_{r}^2\bm{I}_{N_r})$ is the AWGN at BS~B. To maximize the radar SINR, the optimal solution for the filter $\bm{v}$ can be obtained by solving	the minimum variance distortionless response (MVDR) problem \cite{yang2026robust}, i.e., $\bm{v}^*=\beta\left(\bm{\Xi}+\sigma_r^2 \bm{I}_{N_r}\right)^{-1} \bm{A}_{0}(\tilde{\bm{r}},\tilde{\bm{t}})\bm{x}\in \mathbb{C}^{N_r \times 1}$
	where $\beta$ is a non-zero auxiliary constant and $\bm{\Xi}=\sum_{q=1}^{Q}\zeta_{q}^{2}\bm{A}_{q}(\tilde{\bm{r}},\tilde{\bm{t}})\bm{FF}^H\bm{A}_{q}(\tilde{\bm{r}},\tilde{\bm{t}})^{H}\in \mathbb{C}^{N_r \times N_r}$. Then, the average radar SINR can be derived as \cite{yang2026robust}, i.e.,

    \begin{align}
&\Gamma_{r,avr}(\bm{F},\tilde{\bm{r}},\tilde{\bm{t}}) \notag
= \mathbb{E}\left[\frac{\zeta_0^2 \left| \bm{v}^H \bm{A}_0(\tilde{\bm{r}},\tilde{\bm{t}})\bm{x} \right|^2
    }{\bm{v}^H \left( \bm{\Xi}+\sigma_r^2\bm{I}_{N_r} \right)\bm{v}}
    \right] \notag \\
&\qquad \qquad \overset{(a)}{=}
\mathbb{E}\left[
\zeta_0^2\bm{x}^{H}\bm{A}_0(\tilde{\bm{r}},\tilde{\bm{t}})^{H}
\left(\bm{\Xi}+\sigma_r^2\bm{I}_{N_r}\right)^{-1}
\bm{A}_0(\tilde{\bm{r}},\tilde{\bm{t}})
\bm{x}
\right] \notag \\
&\qquad \qquad \overset{(b)}{=}
\operatorname{tr}\left(
\bm{\Phi}\bm{F}\bm{F}^{H}
\right)
\triangleq
\Gamma_r(\bm{F}, \tilde{\bm{r}}, \tilde{\bm{t}}),\label{eq_1}
\end{align}
	where $\bm{\Phi}=\zeta_{0}^{2}\bm{A}_{0}(\tilde{\bm{r}},\tilde{\bm{t}})^{H}(\bm{\Xi}+\sigma_{r}^{2}\bm{I}_{N_r})^{-1}\bm{A}_{0}(\tilde{\bm{r}},\tilde{\bm{t}})\in \mathbb{C}^{N_t \times N_t}$. Note that step (a) follows from the optimal receiving filter $\bm{v}^*$, and step (b) holds due to $\mathbb{E}[\bm{x}\bm{x}^H]=\bm{F}\bm{F}^H$.

	\subsection{Problem Formulation}\par
	Based on the above performance metrics, we focus on the maximization of the radar SINR by jointly designing the beamforming vectors and the positions of the transmit and  receive antennas. Specifically, the optimization problem can be formulated as
    \begin{subequations}
    \label{model_1}
	\begin{align}
		\max_{\bm{F},\tilde{\bm{t}},\tilde{\bm{r}}}  \ &\Gamma_{r}(\bm{F},\tilde{\bm{r}},\tilde{\bm{t}})  \label{model_1a}\\
		\text{s.t. } &\Gamma_{k}\geq\gamma_{k},\ \forall k, \label{model_1b}\\
        &\Gamma_{e,k}\leq\gamma_{e}, \ ||\Delta\bm{h}_{e}||_{2}\leq\varepsilon_{e}, \ \forall k,\label{model_1c}\\
        &t_1 \ge 0,\ t_{N_t} \le D,\ t_n - t_{n-1} \ge d,\ 2 \le n \le N_t, \label{model_1d}\\
        &r_1 \ge 0,\ r_{N_r} \le D,\ r_n - r_{n-1} \ge d,\ 2 \le n \le N_r,\label{model_1e}\\
		&||\bm{F}||_F^2 \le P_t, \label{model_1f}
	\end{align}	     
    \end{subequations}
where $\gamma_k$ represents the required communication SINR threshold, and \eqref{model_1c} ensures that eavesdropping SINR does not exceed the secure transmission SINR threshold $\gamma_e$. In addition, $D$ denotes the feasible movement range for both the transmitting and receiving MAs in the one-dimensional antenna region, $d$ represents the minimum distance between the MAs to prevent coupling effects, and $P_t$ is the maximum transmission power. It is challenging to solve \eqref{model_1} due to the non-concavity of \eqref{model_1a}, the non-convexity of constraints \eqref{model_1b} and \eqref{model_1c}, as
well as the coupling of the variables. 
	\section{Proposed Algorithm}\par 
   In this section, we first reformulate the objective function in \eqref{model_1a} into a more tractable form by using the FP technique. Then, we develop a BCD-based algorithm, the details of which are elaborated as follows.
    \subsection{Problem Reformulation}
 To address the complex fractional objective function in \eqref{model_1a}, we utilize the FP technique in \cite{shen2018fractional} to equivalently transform the original objective function into a tractable form as
\begin{align}
	\nonumber
	&\bar{\Gamma}_r(\bm{F},\tilde{\bm{r}},\tilde{\bm{t}}, \bm{\Lambda}) \\
    &= \zeta_0^2\mathrm{tr}[ 2\Re\left\{\bm{F}^{H}\bm{A}_0(\tilde{\bm{r}},\tilde{\bm{t}})^{H}\bm{\Lambda}\right\} - \bm{\Lambda}^{H}(\bm{\Xi}+\sigma_r^2\bm{I}_{N_r})\bm{\Lambda} ],
    \label{obj_function}
\end{align}
where $\bm{\Lambda} \in \mathbb{C}^{N_r \times M}$ is an auxiliary variable.  
   Then, the BCD algorithm is proposed to solve the problem \eqref{model_1}, the details of which are elaborated as follows.
    \subsection{Updating Auxiliary Variable}\par  
    Note that determining the optimal value of $\bm{\Lambda}$ is an unconstrained optimization problem, and the optimal solution  $\bm{\Lambda}^*$
    can be directly given by
\begin{equation}
			\bm{\Lambda}^{*} = \left(\bm{\Xi}+\sigma_r^2\bm{I}_{N_r}\right)^{-1}\bm{A}_0(\tilde{\bm{r}},\tilde{\bm{t}})\bm{F}.\label{Form_Lambda}
		\end{equation}
	\subsection{Updating Beamforming}\par  
		With $\tilde{\bm{r}}$, $\tilde{ \bm{t}}$, and $\bm{\Lambda}$ being fixed, we focus on the optimization of the transmit beamforming matrix $\bm{F}$.
        We note that the non-convexity of the problem in \eqref{model_1} arises from the quadratic terms associated with the $m$-th beamforming vector, $\bm{f}_m$ within $\bm{F}$. Firstly, the non-convex constraints \eqref{model_1b} and \eqref{model_1c} can be approximated as: \par 
        \begin{align}
      &|\bm{h}_k(\tilde{\bm{t}})^H \bm{f}_k|^2 \geq \gamma_k[\bm{h}_k(\tilde{\bm{t}})^H \bm{F}_{-k} \bm{F}_{-k}^H\bm{h}_k(\tilde{\bm{t}}) + \sigma_k^2],\ \forall k,\label{constrain_1.2}\\
        & ||\bm{h}_e(\tilde{\bm{t}})^H\bm{F}_{-k}||_2^2 \geq \beta_{e,k}-\sigma_e^2,\ ||\Delta\bm{h}_{e}||_{2}\leq\varepsilon_{e},\ \forall k,\label{constrain_2.2b}\\  
        &|\bm{h}_e(\tilde{\bm{t}})^H \bm{f}_k|^2 \leq \gamma_e \beta_{e,k},\ ||\Delta\bm{h}_{e}||_{2}\leq\varepsilon_{e},\ \forall k,\label{constrain_2.2a}
    \end{align}
        where $\bm{F}_{-k}=[\bm{f}_1,\cdots,\bm{f}_{k-1},\bm{f}_{k+1},\cdots,\bm{f}_{M}]\in \mathbb{C}^{N_t\times(M-1)}$, and $\bm{\beta}_e=[\beta_{e,1},\cdots,\beta_{e,K}]^T \geq 0 $ are  auxiliary variables.  
        For the non-convex quadratic terms in \eqref{constrain_1.2}, we employ the first-order Taylor expansion  at the local point $\bm{f}_k^{(n)}$ in \cite{boyd2004convex}. Then the constraint \eqref{constrain_1.2} can be transformed as 
    \begin{align}
    \nonumber
        &|\bm{h}_k(\tilde{\bm{t}})^H\bm{f}_k|^2 \\
        \nonumber
        & \quad \geq 2\Re\left\{\bm{f}_k^{(n),H} \bm{h}_k(\tilde{\bm{t}}) \bm{h}_k(\tilde{\bm{t}})^H \bm{f}_k \right\}- \bm{f}_k^{(n),H} \bm{h}_k(\tilde{\bm{t}}) \bm{h}_k(\tilde{\bm{t}})^H \bm{f}_k^{(n)} \\
        & \quad \geq  \gamma_k[\bm{h}_k(\tilde{\bm{t}})^H \bm{F}_{-k} \bm{F}_{-k}^H\bm{h}_k(\tilde{\bm{t}}) + \sigma_k^2],\ \forall k. \label{constrain_1.2.2}
    \end{align}
Similarly, with the aid of $\bm{h}_e(\tilde{\bm{t}})=\hat{\bm{h}}_e(\tilde{\bm{t}})+\Delta\bm{h}_e$ and first-order Taylor expansion, the inequality \eqref{constrain_2.2b} is reformulated as

        \begin{align}
        \nonumber
            &\Delta \bm{h}_e^H\bm{X}_{e,k}\Delta \bm{h}_e+2\Re\left\{\hat{\bm{h}}_e(\tilde{\bm{t}})^H\bm{X}_{e,k}\Delta\bm{h}_e\right\}+d_{e,k} \geq 0,\\
            &||\Delta\bm{h}_{e}||_{2}\leq\varepsilon_{e},\ \forall k,
        \end{align}
        where $\bm{X}_{e,k}=\bm{F}_{-k}^{(n)}\bm{F}_{-k}^H + \bm{F}_{-k}\bm{F}_{-k}^{H,(n)}-\bm{F}_{-k}^{(n)}\bm{F}_{-k}^{H,(n)}\in \mathbb{C}^{N_t\times N_t}$ and $d_{e,k}= \sigma_e^2-\beta_{e,k}+\hat{\bm{h}}_e(\tilde{\bm{t}})^H\bm{X}_{e,k}\hat{\bm{h}}_e(\tilde{\bm{t}})$. In order to tackle the CSI uncertainty, the S-Lemma in \cite{yang2026robust} is used to transform the above inequality into an equivalent linear matrix inequality (LMI) as
    \begin{equation}
    \begin{bmatrix}
         \bm{X}_{e,k}+\ \omega_{e,k} \bm{I}_{N_t}  & \left( \hat{\bm{h}}_e(\tilde{\bm{t}})^H\bm{X}_{e,k} \right)^H \\
        \hat{\bm{h}}_e(\tilde{\bm{t}})^H\bm{X}_{e,k} & d_{e,k}-\omega_{e,k}\varepsilon_{e}^2
    \end{bmatrix} \succeq \bm{0}_{N_t+1},\ \forall k,
    \label{constrain_2.2b_1}
    \end{equation}
    where $\bm{\omega}_e=[\omega_{e,1},\cdots,\omega_{e,K}]^T \geq 0$ are slack variables. Now, we consider the uncertainty in $\Delta\bm{h}_e$ of \eqref{constrain_2.2a}. We adopt Schur's complement in \cite{boyd2004convex} to equivalently recast \eqref{constrain_2.2a} as
  \begin{equation}
    \begin{bmatrix}
       \gamma_e \beta_{e,k} & t_{e,k} \\
        t_{e,k}^H & 1
    \end{bmatrix} \succeq \bm{0}_2, \ \|\Delta\bm{h}_e\|_2 \leq \varepsilon_e,\ \forall k,\label{eq_2}
    \end{equation}
    where $t_{e,k} = \bm{h}_e(\tilde{\bm{t}})^H\bm{f}_{k}$. Then, by using Nemirovski's lemma in \cite{eldar2005robust} and introducing the slack variables $\bm{\xi}_e=[\xi_{e,1},\cdots,\xi_{e,K}]^T \geq 0$, constraint \eqref{eq_2} can be rewritten as
\begin{equation}
    \begin{bmatrix}
        \gamma_e \beta_{e,k} - \xi_{e,k} & \hat{t}_{e,k} & \bm{0}_{1 \times N_t} \\
        \hat{t}_{e,k}^H & 1 & \varepsilon_e \bm{f}_{k}^H \\
        \bm{0}_{N_t\times 1} & \varepsilon_e \bm{f}_{k} & \xi_{e,k}\bm{I}_{N_t}
    \end{bmatrix} \succeq \bm{0}_{N_t+2}, \ \forall k,\label{constrain_2.2a_1}
    \end{equation}
    where $\hat{t}_{e,k} = \hat{\bm{h}}_e(\tilde{\bm{t}})^H\bm{f}_{k}$. Thus, combining  \eqref{obj_function}, \eqref{constrain_1.2.2}, \eqref{constrain_2.2b_1}, and \eqref{constrain_2.2a_1}, we obtain the approximated optimization problem as 
 \begin{subequations}
    \label{model_2}
	\begin{align}
		\max_{\bm{F},\bm{\beta_{e}},\bm{\omega_{e}},\bm{\xi_{e}}}  & \bar{\Gamma}_r(\bm{F}) \\	     
        \text{s.t.} \quad&  \beta_{e,k}\geq 0,\ \omega_{e,k}\geq 0,\ \xi_{e,k}\geq 0,\ \forall k,     \\
        \nonumber
        & 2\Re\left\{\bm{f}_k^{(n),H} \bm{h}_k(\tilde{\bm{t}}) \bm{h}_k(\tilde{\bm{t}})^H \bm{f}_k \right\}- |\bm{f}_k^{(n),H} \bm{h}_k(\tilde{\bm{t}})|^2 \\
        & \quad \geq  \gamma_k[\bm{h}_k(\tilde{\bm{t}})^H \bm{F}_{-k} \bm{F}_{-k}^H\bm{h}_k(\tilde{\bm{t}}) + \sigma_k^2],\ \forall k\\
        \nonumber
        &\eqref{model_1f},\  \eqref{constrain_2.2b_1},\ \eqref{constrain_2.2a_1}, 
        \end{align}
\end{subequations}
which is convex and can be solved by CVX tool \cite{boyd2004convex}.

\subsection{Updating Antenna Positions}\par 
In this subsection, we focus on the robust design for the transmit MA positions, $\tilde{\bm{t}}$, as the procedure for the receive MA positions, $\tilde{\bm{r}}$, follows a similar methodology and is omitted for brevity.

To decouple $\tilde{\bm{t}}$ in \eqref{obj_function}, the objective function $\bar{\Gamma}_r(\tilde{\bm{t}})$ is first transformed into a more tractable form. Note that the terms independent of $\tilde{\bm{t}}$ are omitted. Letting $\bm{b} = \zeta_0^2\bm{F}\bm{\Lambda}^H\bm{a}_{s,r}(\tilde{\bm{r}})$, $c_q = \zeta_0^2\zeta_q^2\bm{a}_{q,r}(\tilde{\bm{r}})^H\bm{\Lambda}\bm{\Lambda}^H\bm{a}_{q,r}(\tilde{\bm{r}})$ and $\bm{D}_F = \bm{F}\bm{F}^H$, $\bar{\Gamma}_r(\tilde{\bm{t}})$ can be reformulated as
\begin{equation}
\bar{\Gamma}_r(\tilde{\bm{t}}) = 2\Re\left\{ \bm{b}^H \bm{a}_{s,t}(\tilde{\bm{t}})\right\} 
    - \sum_{q=1}^{Q} c_q \bm{a}_{q,t}(\tilde{\bm{t}})^H \bm{D}_F \bm{a}_{q,t}(\tilde{\bm{t}}).
\end{equation}

Thus, the problem in \eqref{model_1} is recast as
\begin{subequations}
\label{model_3}
	\begin{align}
        \max_{\tilde{\bm{t}}} \quad & \bar{\Gamma}_r(\tilde{\bm{t}})  \label{model_3a}\\
		\text{s.t. } &\Gamma_{k}(\tilde{\bm{t}})\geq\gamma_{k},\ \forall k,\label{model_3b}\\
        &\Gamma_{e,k}(\tilde{\bm{t}})\leq\gamma_{e},\ ||\Delta\bm{h}_{e}||_{2}\leq\varepsilon_{e},\ \forall k, \ \eqref{model_1d}.\label{model_3c}
	\end{align}	
    \end{subequations}
The problem \eqref{model_3} is intractable due to the non-concavity of the objective function, the non-convexity of the constraints. To address these issues, the SCA technique can be applied in \cite{boyd2004convex}. We first address the non-concavity of the objective function $\bar{\Gamma}_r(\tilde{\bm{t}})$ in \eqref{model_3a}. Specifically, according to the second-order Taylor expansion theorem in \cite{boyd2004convex},  $\bar{\Gamma}_r(\tilde{\bm{t}})$ can be lower bounded by a quadratic surrogate concave function, i.e.,
\begin{equation}
\bar{\Gamma}_r(\tilde{\bm{t}}) \geq \bar{\Gamma}_r(\tilde{\bm{t}}^{(n)}) + \nabla \bar{\Gamma}_r(\tilde{\bm{t}}^{(n)})^T (\tilde{\bm{t}} - \tilde{\bm{t}}^{(n)}) - \frac{\delta_t}{2} ||\tilde{\bm{t}} - \tilde{\bm{t}}^{(n)}||_2^2,\label{eq_39}
\end{equation}
where $\tilde{\bm{t}}^{(n)}$ denotes the positions of the MAs obtained in the $n$-th iteration. $\nabla \bar{\Gamma}_r(\tilde{\bm{t}}^{(n)}) $ is the gradient vector at $\tilde{\bm{t}}^{(n)}$, and $\delta_t$ is a positive real number that satisfies $\delta_t \bm{I}_{N_t}\succeq  \nabla^2 \bar{\Gamma}_r(\tilde{\bm{t}})$ , which can be  performed in a similar fashion as \cite{yang2026robust}. Similarly, the constraints in \eqref{model_3b} can be first recast as
\begin{equation}
\underbrace{\bm{h}_k(\tilde{\bm{t}})^H \bm{R}_k \bm{h}_k(\tilde{\bm{t}})}_{G_k(\tilde{\bm{t}})} + \gamma_k \sigma_k^2 \leq 0,\ \forall k,
\end{equation}
where $\bm{R}_k = \sum_{u\neq k}^{M} \gamma_k \bm{f}_u \bm{f}_u^H - \bm{f}_k \bm{f}_k^H \in \mathbb{C}^{N_t \times N_t}$. As $G_k(\tilde{\bm{t}})$ is neither convex nor concave with respect to (w.r.t.) $\tilde{\bm{t}}$, we construct a surrogate function that serves as an upper bound of $G_k(\tilde{\bm{t}})$ based on the second-order Taylor expansion, i.e.,
\begin{equation}
G_k(\tilde{\bm{t}}) \leq G_k(\tilde{\bm{t}}^{(n)}) + \nabla G_k(\tilde{\bm{t}}^{(n)})^T (\tilde{\bm{t}} - \tilde{\bm{t}}^{(n)}) + \frac{\delta_k}{2} ||\tilde{\bm{t}} - \tilde{\bm{t}}^{(n)}||_2^2,\label{eq_43}
\end{equation}
where $\nabla G_k(\tilde{\bm{t}}^{(n)})$ denotes the gradient vector over $\tilde{\bm{t}}^{(n)}$. A positive real number $\delta_k$ is selected to satisfy $\delta_k \bm{I}_{N_t}  \succeq\nabla^2 G_k(\tilde{\bm{t}}) $. Similarly, the constraint \eqref{model_3c} can be recast as 
  \begin{equation}
\underbrace{\bm{h}_e(\tilde{\bm{t}})^H \bm{R}_{e,k} \bm{h}_e(\tilde{\bm{t}})}_{G_{e,k}(\tilde{\bm{t}})} + \gamma_e \sigma_e^2 \geq 0, \label{eq_45}
\end{equation}
    where $\bm{R}_{e,k} = \sum_{u\neq k}^{M} \gamma_e \bm{f}_u \bm{f}_u^H - \bm{f}_k \bm{f}_k^H $.
    With the aid of $\bm{h}_e(\tilde{\bm{t}}) = \hat{\bm{h}}_e(\tilde{\bm{t}}) + \Delta \bm{h}_e$,  \eqref{eq_45} can be reformulated as
   \begin{align}
\nonumber
&\Delta \bm{h}_{e}^H\bm{R}_{e,k} \Delta\bm{h}_{e}+
\underbrace{2\Re\left\{
\Delta\bm{h}_{e}^H\bm{R}_{e,k}\hat{\bm{h}}_{e}(\tilde{\bm{t}})\right\}
}_{\triangleq \tilde{A}_{e,k}(\tilde{\bm{t}})}\\
&\qquad \qquad \qquad \qquad+
\underbrace{\hat{\bm{h}}_{e}^H(\tilde{\bm{t}})
\bm{R}_{e,k}\hat{\bm{h}}_{e}(\tilde{\bm{t}})}_{\triangleq \tilde{B}_{e,k}(\tilde{\bm{t}})}+\gamma_{e}\sigma_e^{2}\geq 0.
 \label{eq_4}
\end{align}
Note that $\tilde{A}_{e,k}(\tilde{\bm{t}})$
and $\tilde{B}_{e,k}(\tilde{\bm{t}})$
are neither convex nor concave w.r.t. $\tilde{\bm{t}}$, and thus
we apply the second-order Taylor expansion theorem to approximate the original function with a concave function
w.r.t. $\tilde{\bm{t}}$ as
\begin{equation}
\left\{
\begin{aligned}
&\Delta\bm{h}_{e}^H\bm{R}_{e,k}\Delta\bm{h}_{e}+
2\Re\left\{\Delta\bm{h}_{e}^{H}\bm{U}_{e,k}(\tilde{\bm{t}})\right\}+\tilde{d}_{e,k}(\tilde{\bm{t}},\tilde{\bm{t}}^{(n)})\geq 0,
\\
&\varepsilon_{e}^2-\Delta\bm{h}_{e}^{H}\Delta\bm{h}_{e}\geq 0,
\end{aligned}\right.\label{eq_5}
\end{equation}
where $\bm{U}_{e,k}(\tilde{\bm{t}})=\bm{R}_{e,k} \hat{\bm{h}}_e(\tilde{\bm{t}}^{(n)})+\bm{S}_{e,k}(\tilde{\bm{t}}^{(n)})\big(\tilde{\bm{t}}-\tilde{\bm{t}}^{(n)}\big) \in \mathbb{C}^{N_t \times 1}$, $ \bm{S}_{e,k}(\tilde{\bm{t}})=[\bm{s}_{e,k}(t_1),\cdots,\bm{s}_{e,k}(t_{N_t})]\in \mathbb{C}^{N_t \times N_t}$, and\\ $\bm{s}_{e,k}(t_n)$$ =j \frac{2\pi}{\lambda}\sqrt{\frac{N_t}{L_e}}\,R_{e,k}(:,n)\sum_{l=1}^{L_e} \cos\theta_{e,l}\hat{\rho}_{e,l}e^{j\frac{2\pi}{\lambda} t_n \cos\theta_{e,l}} \in \mathbb{C}^{N_t \times 1}$. Here, $ \tilde{d}_{e,k}(\tilde{\bm{t}},\tilde{\bm{t}}^{(n)})=-\frac{1}{2}(\delta_{A,k}+\delta_{B,k})\left\|\tilde{\bm{t}}-\tilde{\bm{t}}^{(n)}\right\|_2^2+ \tilde{B}_{e,k}\left(\tilde{\bm{t}}^{(n)}\right) + \nabla \tilde{B}_{e,k}\left(\tilde{\bm{t}}^{(n)}\right)^T \left(\tilde{\bm{t}}-\tilde{\bm{t}}^{(n)}\right) +\gamma_e \sigma_e^2$, where $\delta_{A,k}\bm{I}_{N_t} \succeq \nabla^2\tilde{A}_{e,k}(\tilde{\bm{t}})\ \text{and }\delta_{B,k}\bm{I}_{N_t}\succeq \nabla^2\tilde{B}_{e,k}(\tilde{\bm{t}})$. Based on the S-Lemma, the  inequality in \eqref{eq_5} can be further given by
\begin{equation}
\begin{bmatrix}
\Omega_{e,k}\bm{I}_{N_t}+\bm{R}_{e,k}
&\bm{U}_{e,k}(\tilde{\bm{t}})
\\[0.5ex]
\bm{U}_{e,k}(\tilde{\bm{t}})^H
&\tilde{d}_{e,k}(\tilde{\bm{t}},\tilde{\bm{t}}^{(n)})-\Omega_{e,k}\varepsilon_{e}^2
\end{bmatrix}
\succeq \bm{0}_{N_t+1},\label{eq_6}
\end{equation}
where  $\bm{\Omega}_e=[\Omega_{e,1},\cdots,\Omega_{e,K}]^T\geq 0$ are  auxiliary variables. However, the quadratic term in $\tilde{d}_{e,k}(\tilde{\bm{t}},\tilde{\bm{t}}^{(n)})$ leads to the non-convexity of \eqref{eq_6}. Then, constraint \eqref{eq_6} can be approximated as  
\begin{equation}
\begin{bmatrix}
\Omega_{e,k}\bm{I}_{N_t}+\bm{R}_{e,k}
&\bm{U}_{e,k}(\tilde{\bm{t}})
\\
\bm{U}_{e,k}(\tilde{\bm{t}})^H
&
\iota_{e,k}
\end{bmatrix}
\succeq \bm{0}_{N_t+1},\label{eq_50}
\end{equation}
where $\bm{\iota_{e}}=[\iota_{e,1},\cdots,\iota_{e,K}]^T$ are auxiliary variables satisfying
$\iota_{e,k}\leq \tilde{d}_{e,k}(\tilde{\bm{t}},\tilde{\bm{t}}^{(n)})-\Omega_{e,k}\varepsilon_{e}^2$ \cite{yang2026robust}. Thus, combining \eqref{eq_39}, \eqref{eq_43}, and  \eqref{eq_50},  we obtain the approximate optimization problem as 
\begin{subequations}
\label{model_4}
\begin{align}
\max_{\tilde{\bm{t}},\bm{\Omega_{e}},\bm{\iota_{e}}} \ & \bar{\Gamma}_r(\tilde{\bm{t}}^{(n)}) + \nabla \bar{\Gamma}_r(\tilde{\bm{t}}^{(n)})^T (\tilde{\bm{t}} - \tilde{\bm{t}}^{(n)}) - \frac{\delta_t}{2} ||\tilde{\bm{t}} - \tilde{\bm{t}}^{(n)}||_2^2   \\
\text{s.t.} \quad & \iota_{e,k}\leq \tilde{d}_{e,k}(\tilde{\bm{t}},\tilde{\bm{t}}^{(n)})-\Omega_{e,k}\varepsilon_{e}^2, \ \Omega_{e,k} \geq 0,\ \forall k,\\
\nonumber
&G_k(\tilde{\bm{t}}^{(n)}) + \nabla G_k(\tilde{\bm{t}}^{(n)})^T (\tilde{\bm{t}} - \tilde{\bm{t}}^{(n)}) \\&\qquad+\frac{\delta_k}{2} ||\tilde{\bm{t}} - \tilde{\bm{t}}^{(n)}||_2^2
+\gamma_k\sigma_k^2 \leq 0,\ \forall k,\\
\nonumber
& \eqref{model_1d},\ \eqref{eq_50},
\end{align}   
\end{subequations}
which is  convex  and can be solved by the CVX tool~\cite{boyd2004convex}.
\subsection{Complexity Analysis}
\begin{algorithm}[t]
\caption{BCD Optimization Algorithm for \eqref{model_1}}
\begin{algorithmic}[1]
\STATE \textbf{Initialize:} $\bm{F}^{[\psi]}, \tilde{\bm{t}}^{[\psi]}, \tilde{\bm{r}}^{[\psi]}, \bm{\Lambda}^{[\psi]}$, set $\psi = 0$.
\REPEAT
        \STATE Update $\bm{\Lambda}^{[\psi+1]}$ via \eqref{Form_Lambda};
        \STATE Update $\bm{F}^{[\psi+1]}$ by solving \eqref{model_2};
        \STATE Update $\tilde{\bm{t}}^{[\psi+1]}$ by solving \eqref{model_4};
        \STATE Update $\tilde{\bm{r}}^{[\psi+1]}$ in a similar fashion as $\tilde{\bm{t}}^{[\psi+1]}$;
        \STATE  Let $\psi = \psi + 1$;
\UNTIL{The objective value \eqref{obj_function} converges.}
\RETURN $\bm{\Lambda}^*, \bm{F}^*, \tilde{\bm{t}}^*, \text{and}\ \tilde{\bm{r}}^*.$
\end{algorithmic}\label{algorithm1}
\end{algorithm}

The procedures of the proposed BCD algorithm for
solving problem \eqref{model_1} are summarized in Algorithm \ref{algorithm1}. The total computational complexity of Algorithm \ref{algorithm1} mainly arises from the complexity of updating $\bm{F}$ and $\tilde{\bm{t}}$. The complexity order of updating the auxiliary variable $\bm{\Lambda}$  is given by $\mathcal{O}\big(N_r^3\big)$. The complexity order of updating the beamforming matrix $\bm{F}$, transmit antenna positions $\tilde{\bm{t}}$, and receive antenna positions $\tilde{\bm{r}}$ are characterized by $\mathcal{O}(N_t^{6.5} K^{1.5} + N_t^{5.5} K^{2.5} + N_t^{4.5} K^{3.5} + N_t^{3.5} K^{4.5})$, $\mathcal{O}( N_t^{4.5}K^{1.5} + N_t^{3.5}K^{2.5} + N_t^{2.5}K^{3.5}  )$ and $\mathcal{O}(N_r^{3.5})$, respectively.

\section{Numerical Results}
Simulations are carried out to evaluate the proposed scheme. We compare the proposed algorithm with three baseline schemes
1) \textbf{Upper Bound}: Based on the proposed algorithm, the Eve channel is assumed to be perfectly known (i.e., no CSI error is considered), which provides an ideal reference for the robust design.
2) \textbf{Fixed-position antenna (FPA)}: All transmit and receive antennas are fixed at their initial positions with the minimum spacing $\Delta d=\lambda/2$, and only the beamforming matrix is optimized. 
3) \textbf{Greedy antenna selection (GAS)}: The movable regions are discretized into candidate antenna ports with spacing $\Delta d=\lambda/2$, and the antenna positions are greedily selected under the minimum-distance constraint.

In our simulation, BS~A and BS~B are located at (0, 0) m and (40, 0) m, respectively. Unless otherwise specified, the parameters are set as
$N_t=N_r=N=6$, $K=3$, $Q=2$, $L_{\mu}=4$, $\gamma_k=10$ dB, $\gamma_e=0$ dB. Note that $\tilde{c}_k^2=C_0 d_k^{-\kappa}$ represents the large-scale path loss, where $C_0=-30$ dB is the expected average channel power gain at the reference distance of 1 m. The path-loss exponents are $\kappa_{user}=\kappa_{Eve}=2.4$, $\kappa_{target}=\kappa_{clutter}=2.6$.
$\sigma_k^2=\sigma_e^2=\sigma_r^2=-80$ dBm,
$\lambda=0.1$ m, $d=\lambda/2$, $D=10\lambda$, and $P_t=20$ dBW,
with Eve uncertainty radius $\varepsilon_e=3\times10^{-4}$.

\begin{figure}[t]
    \centering
    \subfloat[]{%
        \includegraphics[width=0.48\linewidth]{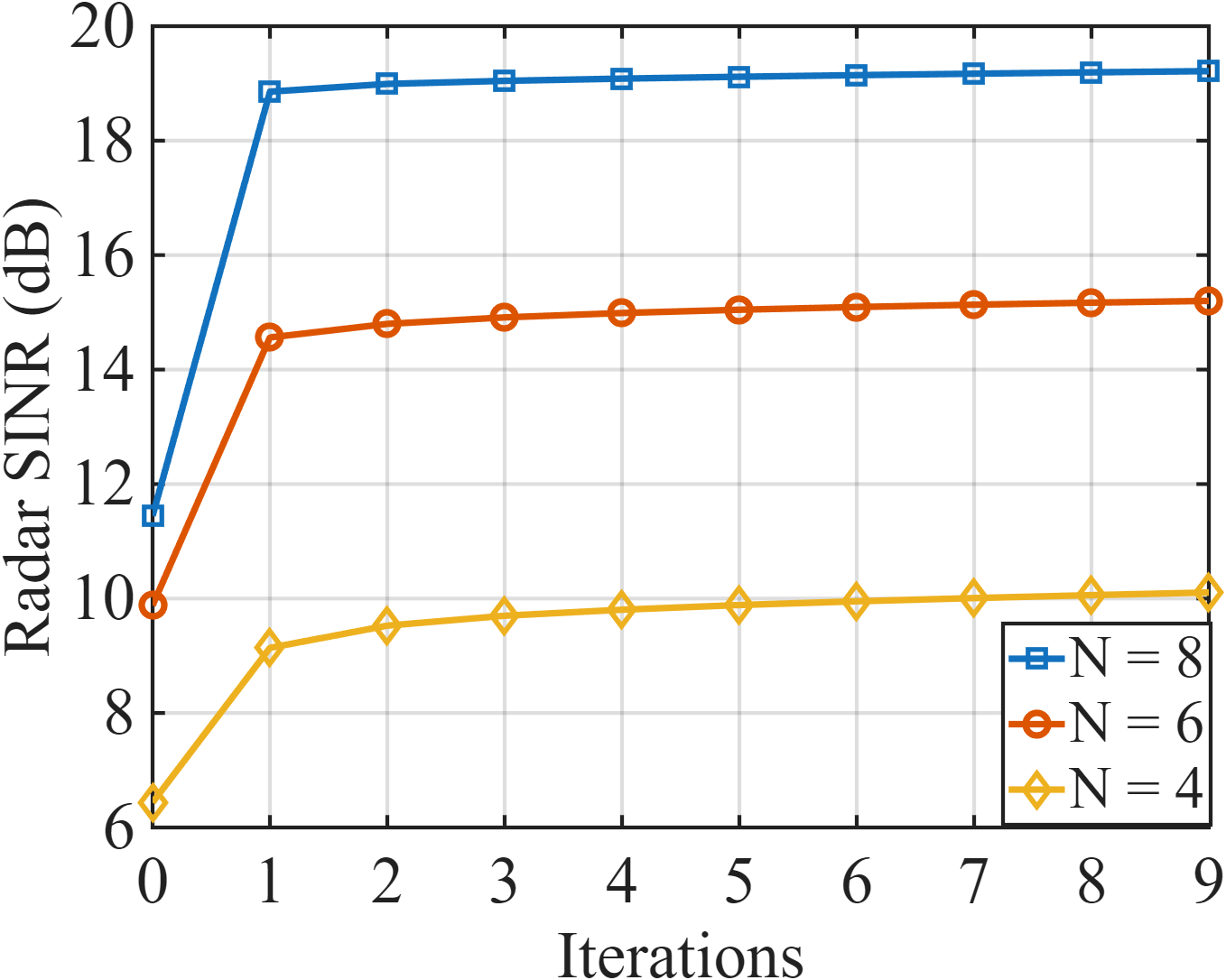}%
        \label{fig:simu_fig2(a)}}
    \hfill
    \subfloat[]{%
        \includegraphics[width=0.48\linewidth]{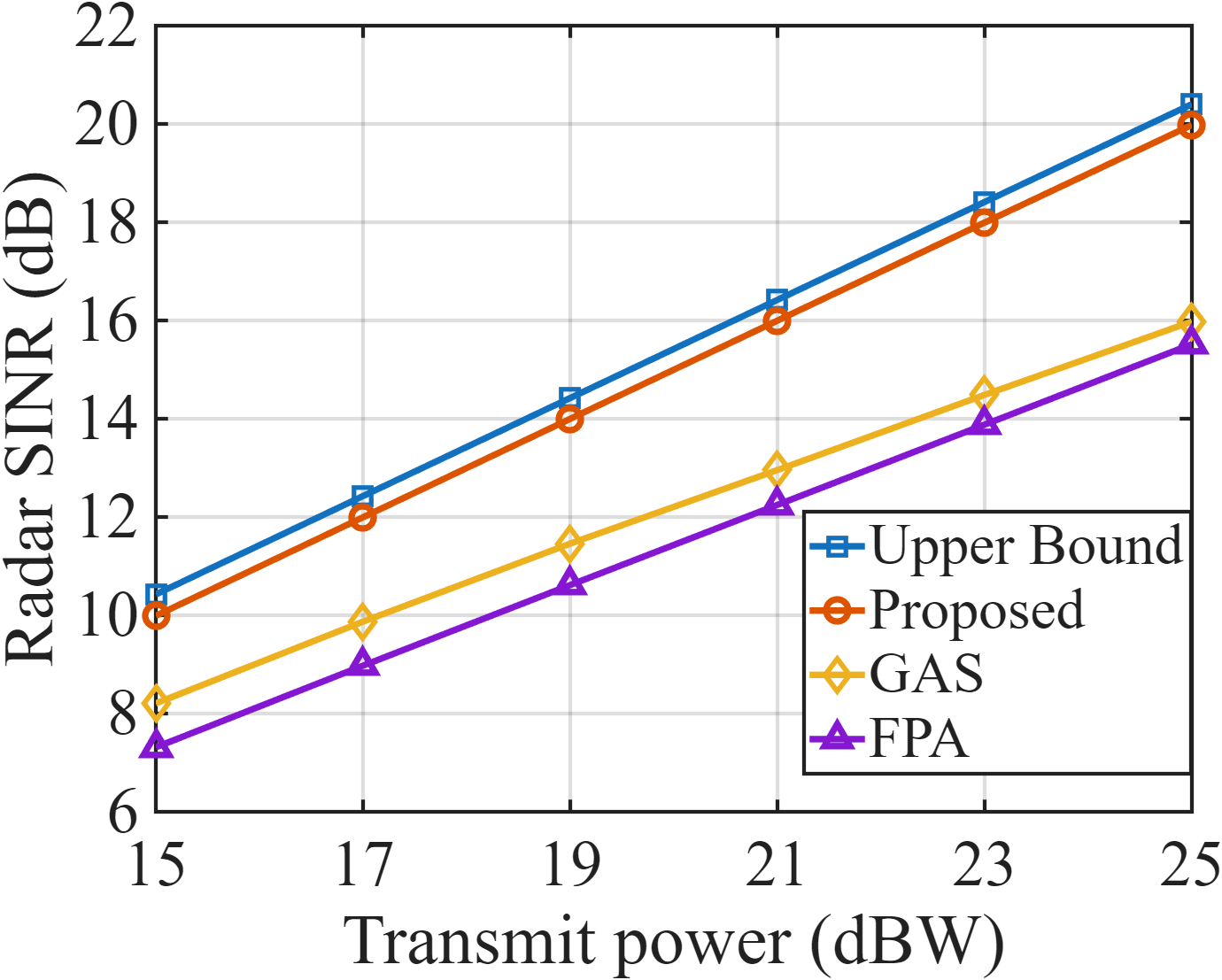}%
        \label{fig:simu_fig2(b)}}
    \caption{(a) Convergence of the proposed algorithm. (b) Radar SINR of different schemes versus transmission power.}
    \label{fig:simu_2}

    \subfloat[]{%
        \includegraphics[width=0.48\linewidth]{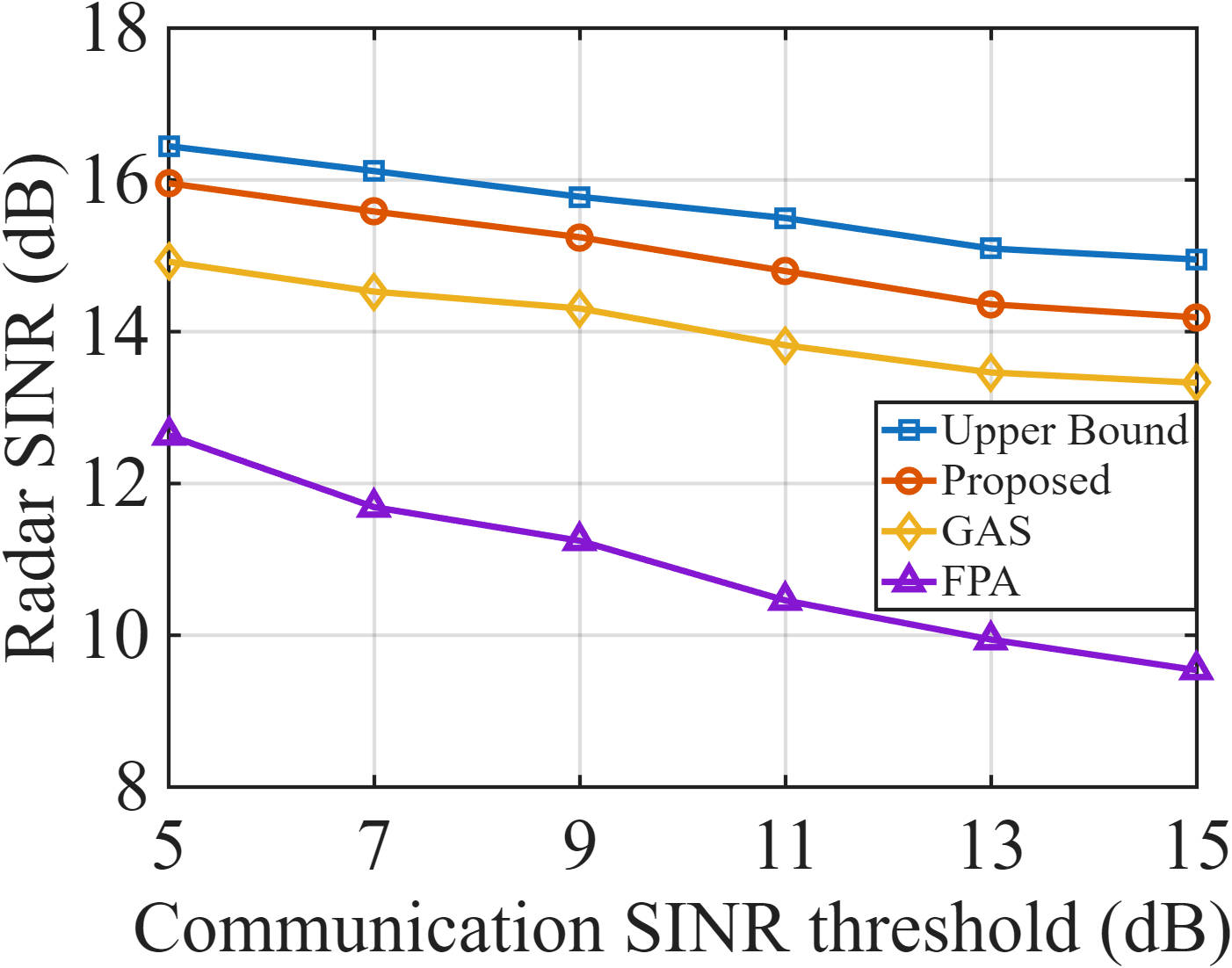}%
        \label{fig:simu_fig3(a)}}
    \hfill
    \subfloat[]{%
        \includegraphics[width=0.48\linewidth]{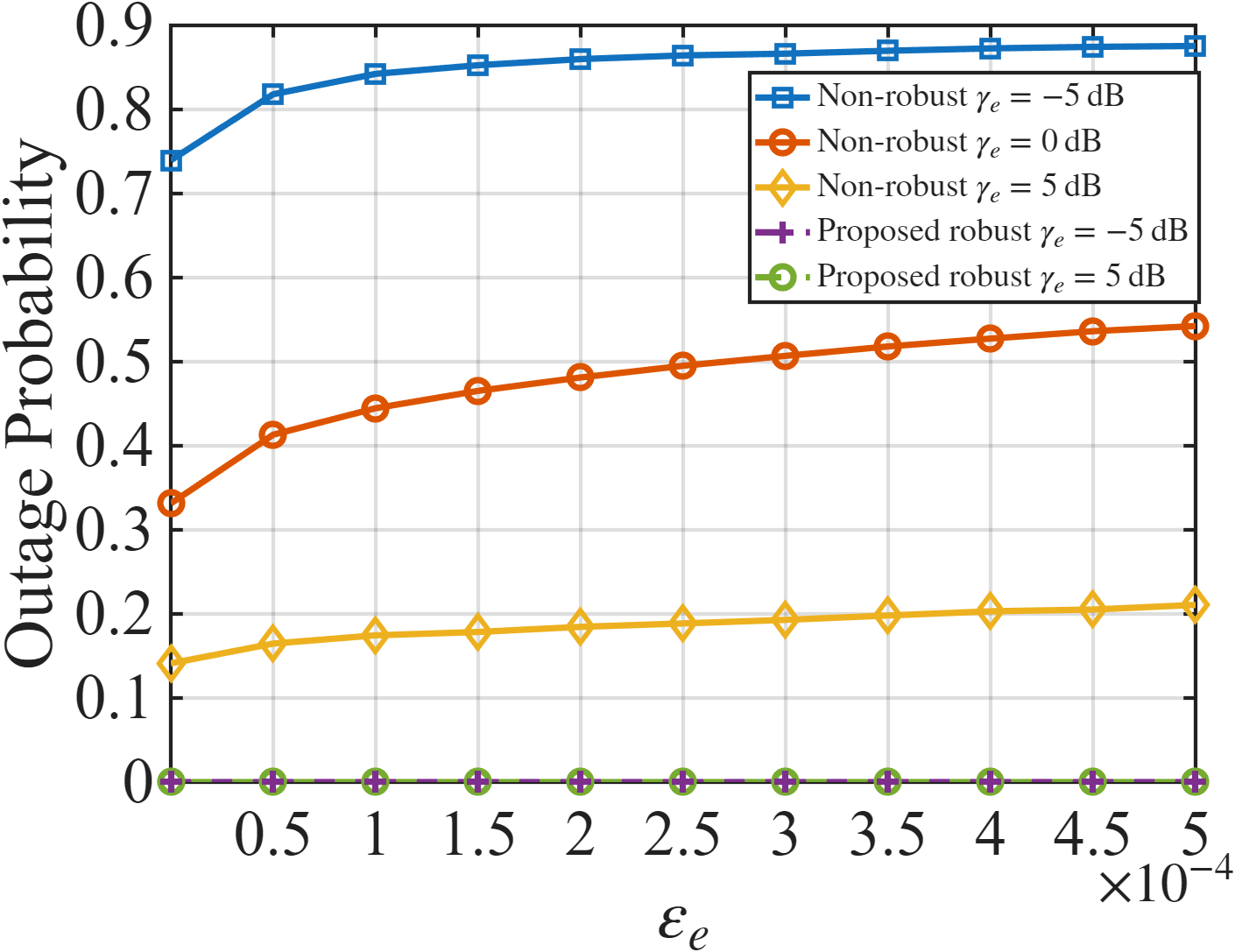}%
        \label{fig:simu_fig3(b)}}
    \caption{(a) Radar SINR versus communication SINR. (b) Outage probability versus the Eve channel uncertainty.}
    \label{fig:simu_3}
\end{figure}

Fig.~\ref{fig:simu_2} shows the convergence behavior and the impact of transmit power. As shown in Fig.~\ref{fig:simu_2}(a), the radar SINR exhibits a monotonic increase with the iteration index and achieves rapid convergence, which verifies the effectiveness of the proposed iterative design. We see that each time we increase two antennas, the radar SINR converges to an additional 4-5 dB. Fig.~\ref{fig:simu_2}(b) shows that the radar SINR of all schemes increases with $P_t$. We highlight that  the proposed scheme consistently outperforms \textbf{FPA} and \textbf{GAS} and remains close to the \textbf{Upper Bound} across the whole power range, which demonstrates that antenna-position optimization can effectively improve sensing performance, bringing it close to the \textbf{Upper Bound}.

Fig.~\ref{fig:simu_3} illustrates the impact of the communication SINR threshold and Eve channel uncertainty. For Fig.~\ref{fig:simu_3}(a), it is observed that the radar SINR decreases as the communication SINR threshold increases, since more resources are required to satisfy the communication SINR constraints. Nevertheless, the proposed scheme remains superior to the benchmark schemes over the whole range, indicating a more favorable sensing-communication tradeoff. Fig.~\ref{fig:simu_3}(b) shows that the outage probability of the non-robust schemes increases with $\varepsilon_e$, whereas the proposed robust scheme maintains zero outage probability over the whole range considered, which demonstrates the robustness of the proposed design against Eve channel uncertainty. In particular, the outage probability is defined as the probability that the Eve SINR constraints $\Gamma_{e,k} \leq \gamma_{e}$ of at least one of them is not satisfied.

\section{Conclusion}
In this letter, we have investigated a robust secure beamforming design for MA-enhanced ISAC systems under imperfect Eve CSI. A radar-SINR maximization problem was formulated by jointly designing the transmit beamforming and the transceiver antenna positions. To solve the intractable problem, we developed a BCD-based iterative algorithm incorporating FP and SCA techniques. Simulation results showed that the proposed technique can significantly improve the radar SINR compared with benchmark schemes. Overall, our work highlights the potential of movable antennas in  ISAC systems, and can find applications in future beyond 6G networks with security considerations.

\end{document}